\documentclass[twocolumn,english,aps,superscriptaddress]{revtex4}
\usepackage[T1]{fontenc}
\usepackage[latin9]{inputenc}
\usepackage{color}
\usepackage{amsmath}
\usepackage{graphicx}
\usepackage{amssymb}

\makeatletter
\@ifundefined{textcolor}{}
{
 \definecolor{BLACK}{gray}{0}
 \definecolor{WHITE}{gray}{1}
 \definecolor{RED}{rgb}{1,0,0}
 \definecolor{GREEN}{rgb}{0,1,0}
 \definecolor{BLUE}{rgb}{0,0,1}
 \definecolor{CYAN}{cmyk}{1,0,0,0}
 \definecolor{MAGENTA}{cmyk}{0,1,0,0}
 \definecolor{YELLOW}{cmyk}{0,0,1,0}
 }

\@ifundefined{definecolor}{\@ifundefined{definecolor}{\@ifundefined{definecolor}{\@ifundefined{definecolor}
 {\usepackage{color}}{}
}{}}{}}{}

\makeatother

\usepackage{babel}

\makeatother

\usepackage{babel}

\makeatother

\usepackage{babel}

\makeatother

\usepackage{babel}

\begin{document}

\title{Tunable Wigner States with Dipolar Atoms and Molecules}

\author{J.\ C.\ Cremon}

\affiliation{Mathematical Physics, LTH, Lund University, SE-22100 Lund, Sweden}

\author{G.\ M.\ Bruun}

\affiliation{Mathematical Physics, LTH, Lund University, SE-22100 Lund, Sweden}

\affiliation{Department of Physics and Astronomy, University of Aarhus, DK-8000
Aarhus C, Denmark}

\author{S.\ M.\ Reimann}

\affiliation{Mathematical Physics, LTH, Lund University, SE-22100 Lund, Sweden}
\begin{abstract}
We study the few-body physics of trapped atoms or molecules with electric or magnetic
dipole moments aligned by an external field. Using exact numerical
diagonalization appropriate for the strongly correlated regime, as
well as a classical analysis, we show how Wigner localization emerges
with increasing coupling strength. The Wigner states exhibit nontrivial
geometries due to the anisotropy of the interaction. This leads to
transitions between different Wigner states as the tilt angle of the
dipoles with the confining plane is changed. Intriguingly, while the
individual Wigner states are well described by a classical analysis,
the transitions between different Wigner states are strongly affected
by quantum statistics. This can be understood by considering the interplay
between quantum-mechanical and spatial symmetry properties. Finally,
we demonstrate that our results are relevant to experimentally realistic
systems.
\end{abstract}
\maketitle
\begin{figure}
\begin{centering}
\includegraphics[width=1\columnwidth]{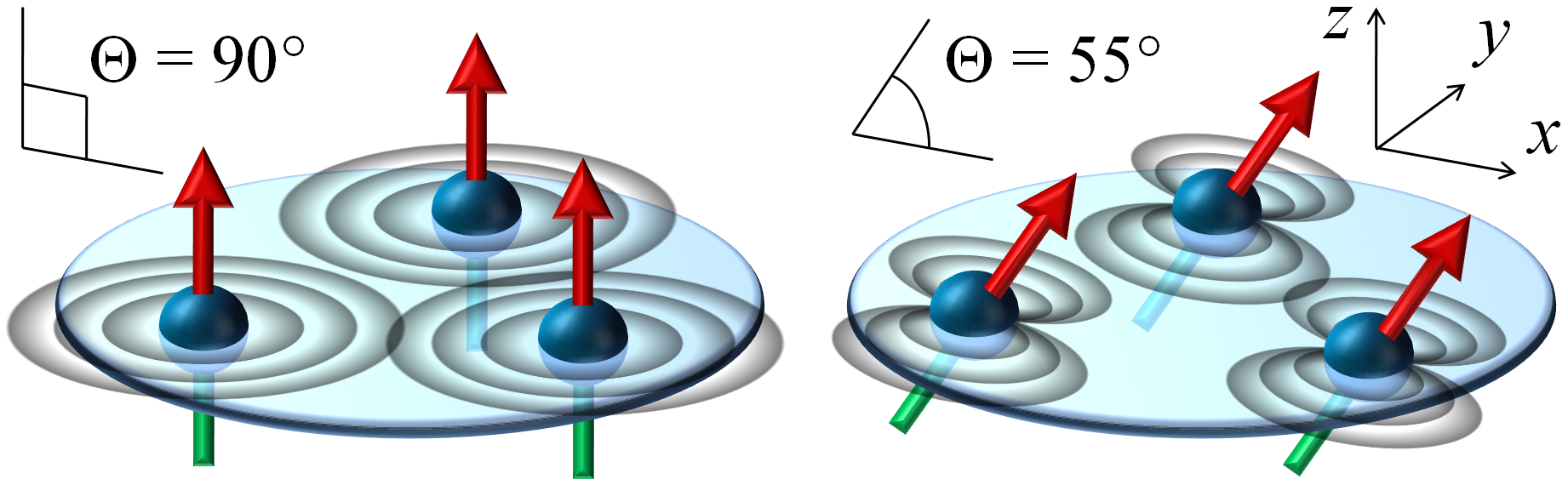} 
\par\end{centering}
\caption{\label{fig:schematic-dipoles}Dipolar particles 
in a quasi-2D trap in the $xy$ plane. The gray circles indicate
the contours of the effective interaction potentials in the plane.
In the left-hand panel the dipoles are perpendicular ($\Theta=90\textdegree$)
to the plane of motion and the interaction is isotropic in the plane,
while in the right-hand panel the dipoles are tilted ($\Theta=55\textdegree$)
and the interaction is anisotropic.}

\end{figure}

Recent experimental advances in cold quantum gases have
placed focus on atoms or molecules with permanent dipole moments. 
A Bose-Einstein condensate of $^{52}$Cr atoms has been realized~\cite{lahaye2007,koch2008}, 
and recently  Dy atoms were cooled and trapped\,\cite{lu2010}.
These atom species have magnetic dipole moments of several
Bohr magnetons. A promising development is the trapping and cooling
of diatomic molecules with electric dipole moments\,\cite{danzl2010,deiglmayr2009,aymar2005}.
The realization of a molecular fermionic $^{40}$K$^{87}$Rb gas was
a significant breakthrough~\cite{KRbJILArefs}.
Dipolar gases offer access to a broad range of novel few- and many-body
physics, in single traps as well
as in optical lattices which has spurred intensive theoretical interest
(see the recent reviews by Baranov \cite{baranov2008rev} and Lahaye
\textit{et al.}~\cite{lahaye2009}). The attractive part of the dipolar interaction leads to a collapse
instability in three dimensions \cite{lahaye2009}. A remedy is to
use traps of reduced dimensionality. For instance, the interaction
between dipoles in a 2D plane is predominantly repulsive when they form a sufficiently large angle with respect  
 to the plane (see Fig. \ref{fig:schematic-dipoles}). This
stabilizes the system against collapse~\cite{bruun2008}.
A variety of interesting many-body states for dipoles in 2D has been
examined theoretically~\cite{dipolesrefs,bruun2008,baranov2008}.

Here, we examine a 2D system of this kind in the regime
of strong repulsive interactions using exact diagonalization as well
as a classical analysis. 
Contrary to the analogous Wigner states of electrons in metals~\cite{wigner1934}

and quantum dots \cite{qdotsrefs}
the anisotropic dipolar interaction is shown to give rise to Wigner
states with nontrivial geometries that depend on the alignment angle.
This leads to transitions between different geometries
that are crucially influenced by quantum statistics. 
We finally argue that our results are experimentally observable.

We consider particles with mass $m$ and a magnetic (electric) dipole
moment $\mu$ ($d$) which is aligned by an external field such that
it lies in the $xz$ plane, forming an angle $\Theta$ with the $x$ axis
(see Fig. \ref{fig:schematic-dipoles}). The interaction between
two electric dipoles separated by a vector $\mathbf{r}$ is \begin{equation}
V(\mathbf{r})=\frac{d^{2}}{4\pi\epsilon_{0}}\frac{1-3\cos^{2}\theta_{rd}}{r^{3}}\label{Vbare}\end{equation}
 where $\theta_{rd}$ is the angle between a dipole moment and $\mathbf{r}$.
The particles are confined in the $xy$-plane by a quasi-two-dimensional
harmonic trap $V_{{\rm trap}}(x,y,z)=m[\omega_{0}^{2}(x^{2}+y^{2})+\omega_{z}^{2}z^{2}]/2$,
with $\omega_{z}\gg\omega_{0}$ so that the particles are in the ground
state orbital of the $z$ direction. The corresponding oscillator
lengths are $l_{z}$ and $l_{0}$ with $l_{z}\ll l_{0}$. Throughout this Letter, we consider  dipoles with no internal spin degrees of freedom. 

The effective interaction $V_{2D}(r,\phi)$ in the $xy$-plane is
obtained by integrating out the harmonic motion in the $z$-direction.
This yields \begin{multline}
\frac{V_{\mathrm{2D}}(r,\phi)}{\hbar\omega_{0}}=\frac{D^{2}}{2\sqrt{2\pi}}\frac{e^{\xi/2}}{(l_{z}/l_{0})^{3}}\biggl((2+2\xi)K_{0}(\xi/2)-2\xi K_{1}(\xi/2)\\
+\cos^{2}\Theta\biggl[-(3+2\xi)K_{0}(\xi/2)+(1+2\xi)K_{1}(\xi/2)\biggl]+\\
+2\cos^{2}\Theta\cos^{2}\phi\biggl[-\xi K_{0}(\xi/2)+(\xi-1)K_{1}(\xi/2)\biggl]\biggl)\label{eq:effective-interaction-function}\end{multline}
 where $\xi=r^{2}/(2\tilde{l}_{z}^{2})$ and $K_{0}$ and $K_{1}$
are irregular modified Bessel functions. 
 (For the special cases $\Theta=90\textdegree$ and $\Theta=0\textdegree$, 
expressions are also given in \cite{yi2006komineas2007,baranov2008}.)
For electric dipoles, $D=\frac{d}{\sqrt{4\pi\epsilon_{0}}}\frac{\sqrt{m}}{\hbar\sqrt{l_{0}}}$
whereas $D=\mu\sqrt{\frac{\mu_{0}}{4\pi}}\frac{\sqrt{m}}{\hbar\sqrt{l_{0}}}$
for magnetic dipoles. The effective interaction (\ref{eq:effective-interaction-function})
reduces to the bare interaction (\ref{Vbare}) when $r\gg l_{z}$.
Depending on the angle $\Theta$, $V_{{\rm 2D}}(r,\phi)$ contains
both attractive and repulsive regions. For $\Theta=90\textdegree$
it is isotropic in the $xy$-plane, whereas it is anisotropic with
increasing attractive regions for decreasing angle. Since the focus
is on Wigner states in this Letter, we restrict our analysis to the
case $\arccos\frac{1}{\sqrt{3}}\approx54.7\textdegree<\Theta\le90\textdegree$,
for which the interaction without the $z$-integration is purely repulsive.

\begin{figure}

\begin{centering}
\includegraphics[width=1\columnwidth]{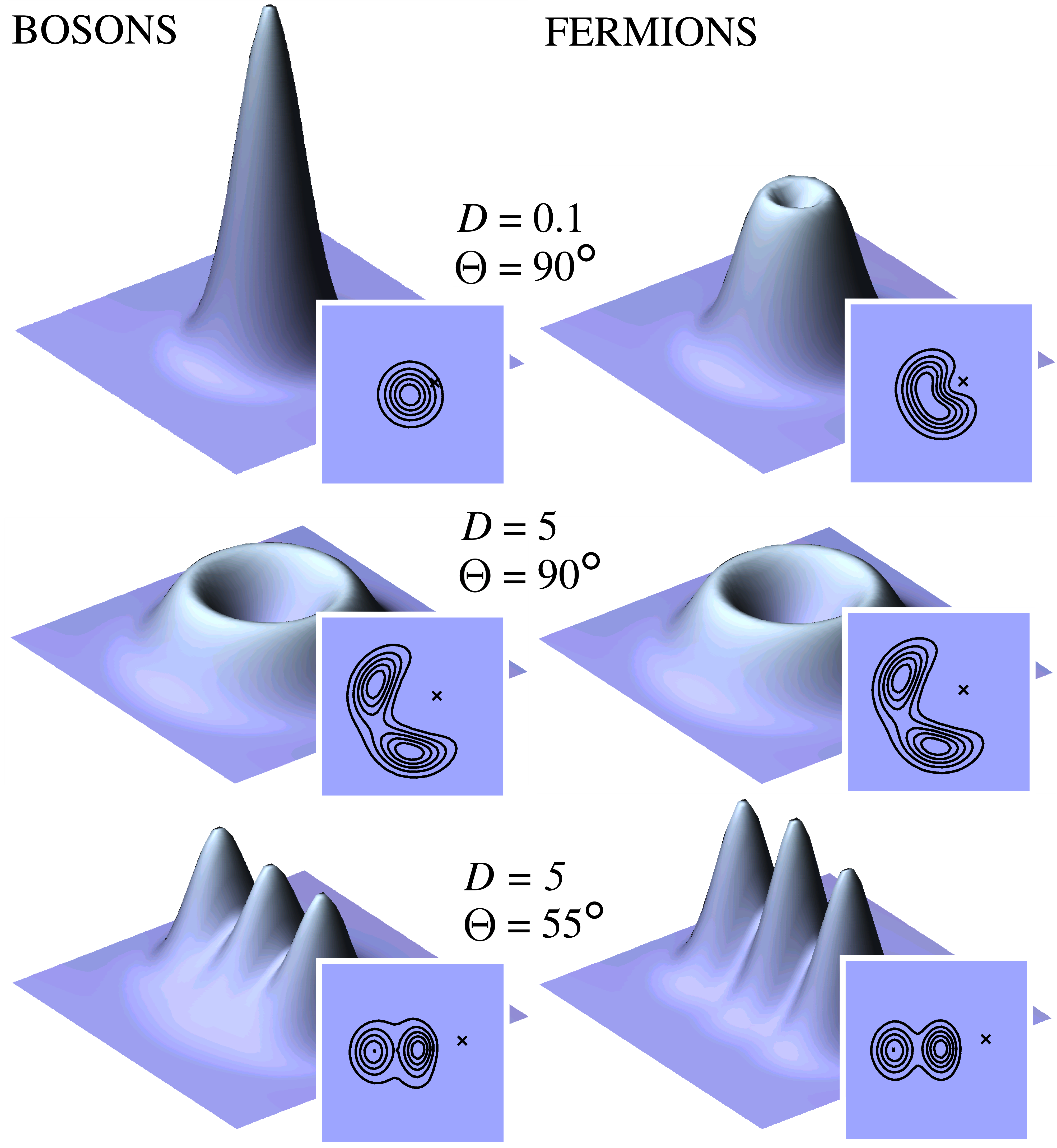} 
\par\end{centering}

\caption{\label{fig:purple-surfaces}
 Particle densities for three different coupling strengths, with dipole
tilt angles $\Theta=90\textdegree$ and $\Theta=55\textdegree$. The
insets show the pair-correlated density functions; the crosses mark
the position of one particle. The $x$ and $y$ axes range from $-4l_{0}$
to $4l_{0}$, both here and in Figs. \ref{fig:crossing-anticrossing}
and \ref{fig:N4-plots}.}

\end{figure}

Our numerical results are obtained by exact diagonalization using
a basis of 2D harmonic oscillator orbitals. The multiparticle basis space is truncated
by including all states with (kinetic and potential) energy up to
some cutoff value $E_{cut}$. Here, good convergence is achieved at
$E_{cut}<20\hbar\omega$, and variations of $E_{cut}$ do not qualitatively
affect our conclusions. Because of the computational effort involved,
the method is limited to particle numbers $N\lesssim4$. In the following
we set $l_{z}/l_{0}=0.1$. Our results are not sensitive to this ratio
as long as it is sufficiently small. In Fig.\ \ref{fig:purple-surfaces}
we plot the particle density of the ground state 
for two different tilt angles $\Theta$ for a system with $N=3$ dipolar
bosons and fermions. For weak coupling, $D=0.1$, the dipoles
essentially form an ideal gas. The dip in the density for the fermions
in Fig.~\ref{fig:purple-surfaces} is simply a shell effect. With
increasing interaction, the dipoles localize in Wigner states. Comparing
the interaction energy with the confinement kinetic energy yields
the condition $D^{2}\gg1$ for Wigner crystallization. Our numerical
results confirm this showing that the crystallized structure emerges
continuously for $D\gtrsim1$ both for bosons and fermions for this
finite-size system; for $D=5$ the localization can be clearly seen
in Fig.\ \ref{fig:purple-surfaces}.

The Hamiltonian and the density of the ground state have rotational
symmetry for $\Theta=90\textdegree$. To demonstrate particle localization,
we therefore study the pair-correlated density 
$\rho({\mathbf{r}},{\mathbf{r}}')=\langle\hat{\psi}^{\dagger}({\mathbf{r}})\hat{\psi}^{\dagger}({\mathbf{r}}')\hat{\psi}({\mathbf{r}}')\hat{\psi}({\mathbf{r}})\rangle$.
Fixing one particle at position ${\mathbf{r}}$, it  gives the probability distribution of the remaining $N-1$ particles. 
We see from Fig.~\ref{fig:purple-surfaces} that for strong interactions
and $\Theta=90\textdegree$, the particles localize at the vertices
of an equilateral triangle. For $\Theta=55\textdegree$, the interaction
is anisotropic in the $xy$-plane and almost vanishing between two
dipoles whose relative vector is parallel to the $x$-axis. Figure
\ref{fig:purple-surfaces} shows that the ground state for $D=5$
features the three particles on a straight line for both bosons and
fermions. This exploits the weaker regions of the interaction and
minimizes the energy. A classical analysis minimizing the trap and
interaction energies of three dipoles in a 2D trap indeed yields that the lowest energy 
configuration is all particles on a line along the $x$-axis for
$\Theta\lesssim62\textdegree$ and the three particles in an equilateral triangle
for $\Theta\gtrsim62\textdegree$.

\begin{figure}
\begin{centering}
\includegraphics[width=1\columnwidth]{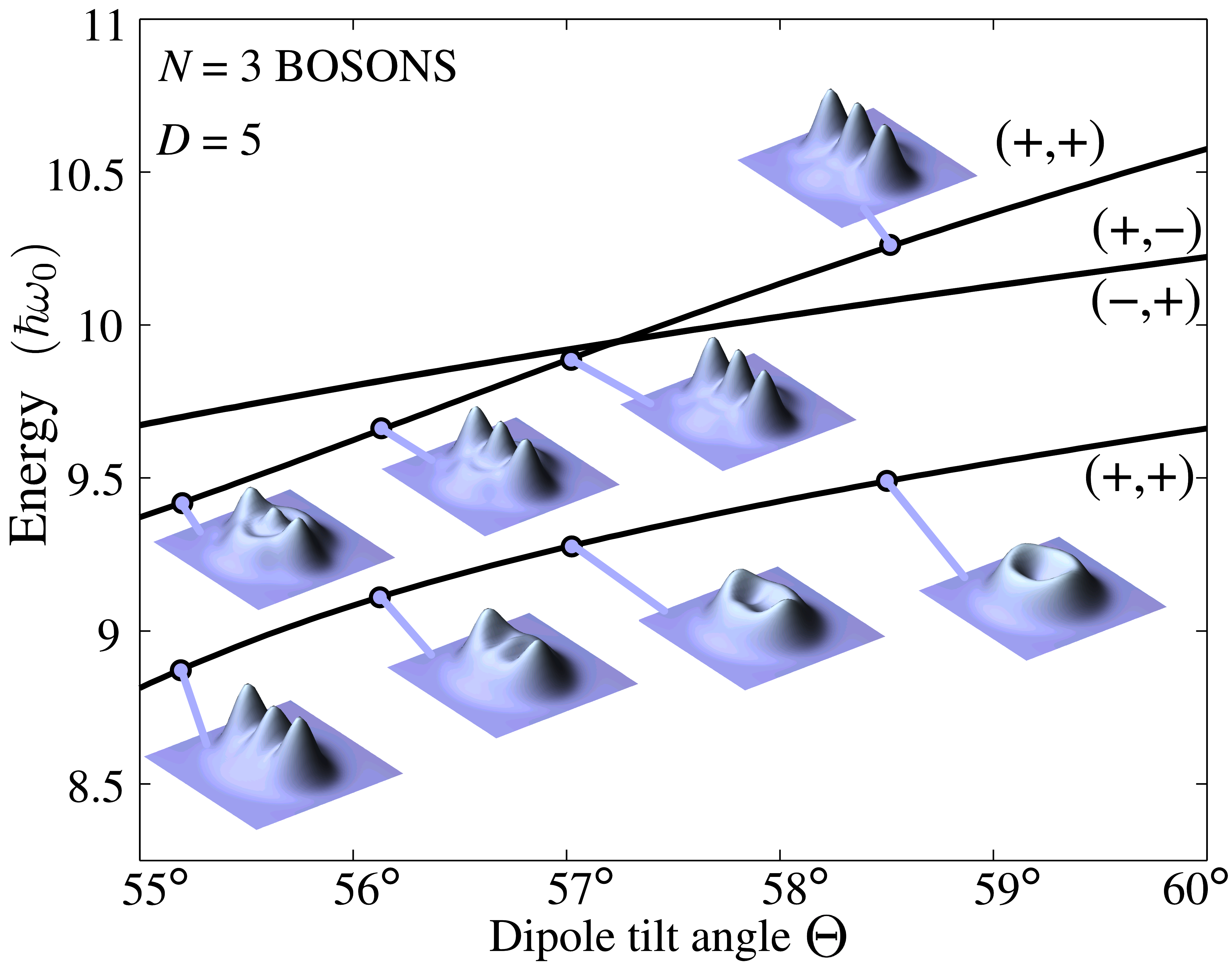}
\includegraphics[width=1\columnwidth]{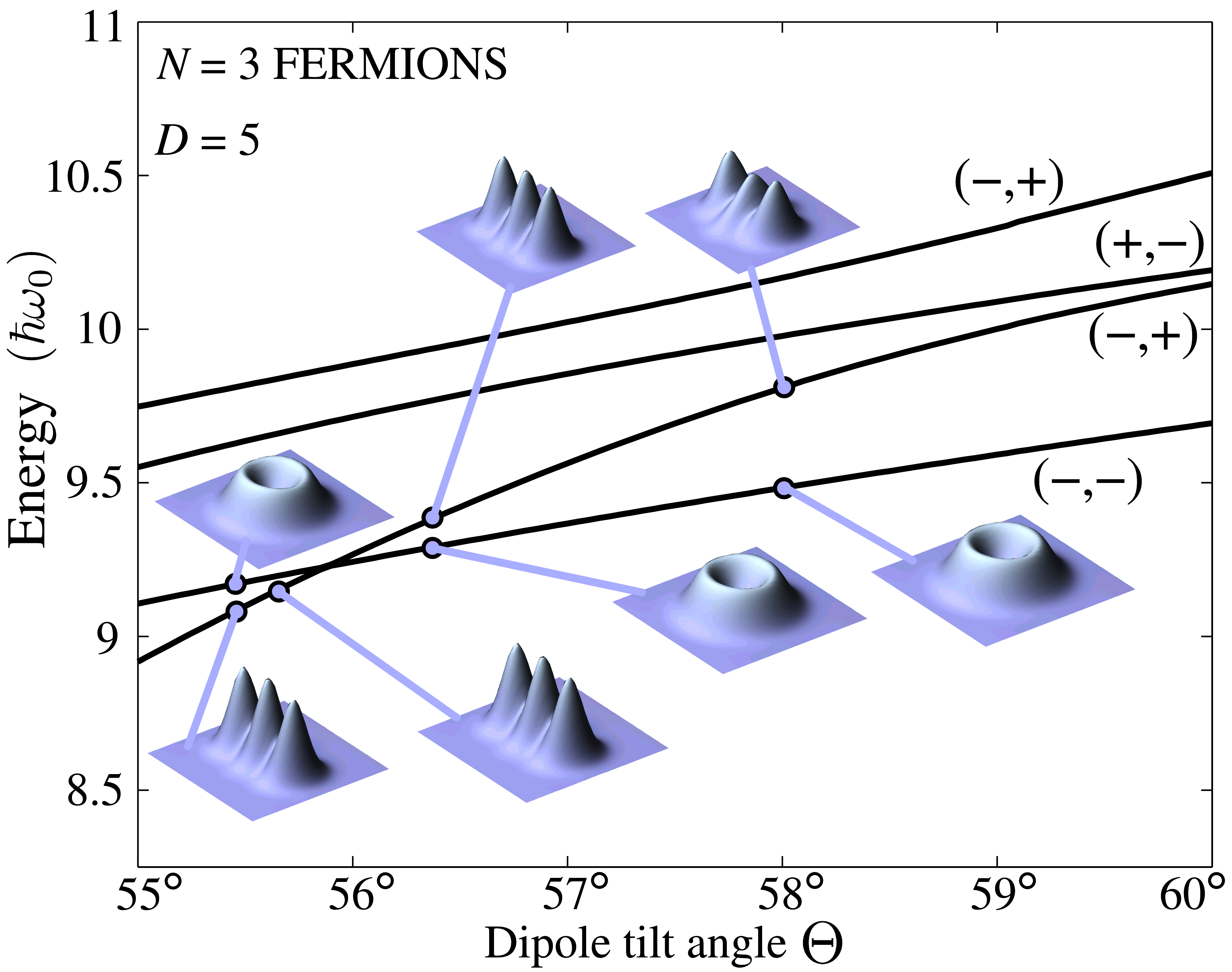} 
\par\end{centering}

\caption{\label{fig:crossing-anticrossing}The four lowest energy states for three strongly
interacting bosonic (top panel) and fermionic (bottom panel) dipoles.
By changing the dipole tilt angle $\Theta$, the ground state changes
from a Wigner state with line geometry to one with triangular
geometry. The signs
in brackets denote $x$- and $y$-parities, as explained in the text.
}

\end{figure}

We now examine the transition from a line shaped to a triangular shaped ground state with increasing angle.
 Figure \ref{fig:crossing-anticrossing}
depicts the lowest energy states of the three dipoles and their density
as a function of the tilt angle $\Theta$. In agreement with the classical
analysis, the ground state changes from a linear to a triangular Wigner
state as $\Theta$ is increased. For $D=5$, the transition occurs
at $\Theta\simeq56\textdegree$ for fermions and $\Theta\simeq57\textdegree$
for bosons which is somewhat lower than the classical prediction $\Theta\simeq62\textdegree$ due to quantum fluctuations. 
 The density of the triangular state appears rotationally
symmetric since the energy of three aligned classical dipoles in an equilateral
triangle can be shown to be independent of its orientation.

There is an intriguing difference between the transitions from 
 linear- to  triangular-shaped Wigner states for fermionic
and bosonic dipoles: For fermions, the transition is a sharp crossing, 
whereas, for bosons, it is a continuous mixing of the triangular and linear
states corresponding to an avoided crossing. This can be understood
by considering that the Hamiltonian is invariant under the mirror
operations $\hat{P}_{x}:$ $(x,y)\rightarrow(-x,y)$ and $\hat{P}_{y}:$
$(x,y)\rightarrow(x,-y)$. The energy eigenstates must therefore also
be eigenstates of $\hat{P}_{x}$ and $\hat{P}_{y}$ with eigenvalues
$\pm1$. The key point is that for the Wigner states, one can infer
their eigenvalues under these symmetry operations from their quantum
statistics. Consider the three-particle line configuration discussed
above. Applying $\hat{P}_{x}$ simply corresponds to exchanging the
coordinates of the two outer localized particles. For bosons, this
must give a plus sign due to the symmetrization of the wave function.
For fermions one must correspondingly get a minus sign. Likewise,
since all three particles have a finite probability to appear on the
$x$-axis, the eigenvalue for $\hat{P}_{y}$ must be $+1$ for both
bosons and fermions. We write this as $(P_{x},P_{y})_{\mathrm{B,line}}=(+,+)$
for the bosonic line state and $(P_{x},P_{y})_{\mathrm{F,line}}=(-,+)$
for the fermionic line state.

The triangular Wigner state is more subtle, as it contains components
with the triangle in all possible orientations. Take the component
with one of the vertices of the triangle lying on the $x$ axis: Operating
with $\hat{P}_{y}$ on this component corresponds to swapping the
particles located at the two other vertices of the triangle symmetrically
placed above and below the $x$-axis; for bosons and fermions this
will give the eigenvalue $+1$ and $-1$ respectively for this component.
Since the Wigner state has to be an eigenfunction of $\hat{P}_{y}$, all other components of the wave function corresponding
to tilted triangles must come in symmetric/antisymmetric pairs mirrored
in the $x$-axis, so that the total wavefunction has the eigenvalue
$+1$ (bosons) or $-1$ (fermions) for $\hat{P}_{y}$. The same analysis
applies to $\hat{P}_{x}$ and we conclude that for the triangular
Wigner state $(P_{x},P_{y})_{\mathrm{B,triangle}}=(+,+)$ for bosons
and $(P_{x},P_{y})_{\mathrm{F,triangle}}=(-,-)$ for fermions. Since
energy levels corresponding to wave functions of the same symmetries
cannot cross, the arguments above explain why the line-triangle transition
for bosons corresponds to an avoided crossing, while it is sharp for
fermions. It is perhaps surprising that quantum statistics plays an
important role for the transitions between essentially classical Wigner
states. This can, however, be understood by noting that the particles
can overlap during the transitions meaning that quantum statistics
matters. 

 \begin{figure}[t]
\begin{center}
\leavevmode

\includegraphics[clip=true,width=1\columnwidth]{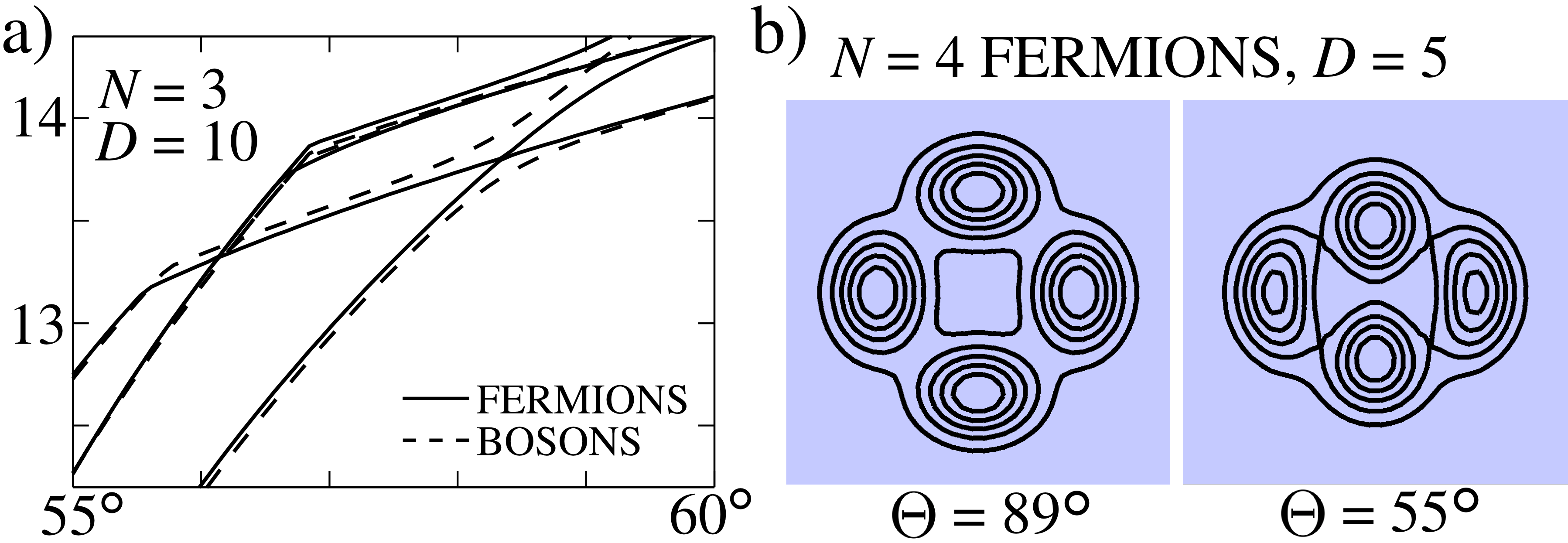}

\caption{(a) The four lowest energy states for very strong interaction (axes as in Fig. 3). 
(b) Density contours for $N=4$ dipolar fermions.}
\label{fig:N4-plots}
\end{center}
\end{figure}

It is illuminating to connect the states in Fig.~\ref{fig:crossing-anticrossing} to the eigenstates of the angular momentum $\hat L_z$
along the $z$-axis for  $\Theta=90\textdegree$: The lowest level 
in Fig.~\ref{fig:crossing-anticrossing} naturally develop into to $L_z=0$ for  $\Theta=90\textdegree$, whereas the next two levels correspond to $L_z\pm3$ and 
not $L_z=\pm 1$ as one would perhaps expect. This is because the strong repulsion favors states with large $|m|$
which are more spatially extended. 

In Fig.~\ref{fig:N4-plots} (a), we plot the four lowest energy states in the limit of very strong interaction $D=10$ (the interaction strength 
scales as $D^2$). Comparing with Fig.~\ref{fig:crossing-anticrossing}, we see  that the  difference between the bosons and fermions decreases with 
increasing coupling. This is as expected since the system approaches classical  behavior in this limit. Reflecting this, the transition regions 
between different ground states where quantum statistics is important are smaller for $D=10$ than for $D=5$: The size of the discussed anticrossing is $\Delta E=0.53$
for $D=5$ and $\Delta E=0.22$ for $D=10$. Also, the transition region has moved to larger angle $\Theta\simeq59\textdegree$ approaching the classical 
value $\Theta\simeq62\textdegree$ with increasing $D$.

\begin{figure}
\begin{centering}
\includegraphics[width=1\columnwidth]{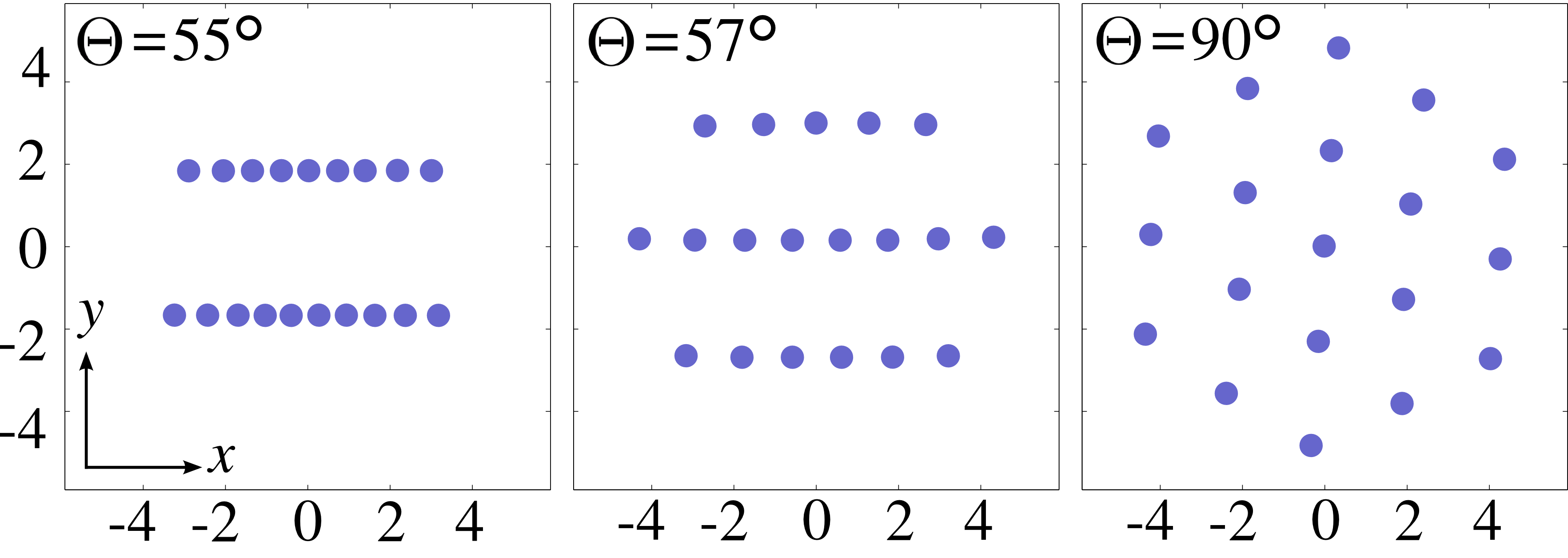} 
\par\end{centering}
\caption{\label{fig:classical-N19}A classical calculation
of the minimum energy configurations of $N=19$ dipoles in a 2D trap.}

\end{figure}

In Fig. \ref{fig:N4-plots} (b), we plot the density of the ground state
of $N=4$ fermionic dipoles for $D=5$. In this case, the dipoles localize in a rhombic
geometry that depends on the tilt angle. To examine the case of higher particle numbers, 
 we present in Fig.~\ref{fig:classical-N19} the result of a classical minimization of the interaction and trap
energy for $N=19$ dipoles. The ground state configurations were found
using an iterative self-consistent procedure adjusting the particle
positions to find the minimal energy. We see that  in analogy with the $N=3,4$ cases considered
above,  the geometry of the lattice depends on the tilt angle. For $\Theta=90\textdegree$,
 the particles  localize in a hexagonal lattice~\cite{buchler2007}. These classical
calculations indicate that our main conclusions for quantum-mechanical
few-body systems should be relevant also for larger systems.

The strong coupling effects considered here are reachable using typical
experimental numbers. The KRb molecule trapped by the JILA group~\cite{KRbJILArefs}
has an electric dipole moment of $0.57$\,Debye whereas the RbCs
molecule studied by the Innsbruck group has an electric dipole moment
of $1.25$\,Debye~\cite{pilch2009,kotochigova2005}. Trapping lengths
of $l_{0}=1$\,$\mathrm{\mu m}$ or $l_{0}=0.1$\,$\mathrm{\mu m}$
yield coupling strengths between $0.6\lesssim D^{2}\lesssim50$. Also,
there are alkali dimers with even larger dipole moments~\cite{aymar2005}. 
Using optical lattices, few-body systems with ultracold molecules may be realized, 
see e.g. the experiment in Ref. \cite{danzl2010}.

To summarize, we studied systems of strongly interacting bosonic and
fermionic dipoles in a 2D harmonic trap, using both exact diagonalization
and classical analysis. The dipolar interaction was shown to lead
to a rich variety of Wigner states with nontrivial geometries which
depend on the tilt angle of the dipoles with respect to the plane.
Even though the Wigner states themselves can be well understood from
a classical analysis, the transitions between different geometries
as the tilt angle is changed depends crucially on the quantum statistics
of the dipoles. We showed how the effects analyzed here are
relevant for typical experimental parameters.

We thank G. Carlsson, J. Grönqvist, O. Karlström and F. Malet 
for helpful comments and suggestions. This
work was financially supported by the Swedish Research Council and
the Kungl. Fysiografiska Sällskapet in Lund.

\end{document}